\documentstyle[12pt,psfig]{article}
\newcommand{\s}{\rm}

\newcommand{\ra}{\rightarrow}
\newcommand{\mn}{\mu \nu}
\newcommand{\be}{\begin{equation}}
\newcommand{\ee}{\end{equation}}

\begin{document}

\begin{flushright}
KUNS-1567.
\end{flushright}
\begin{center}
{\Large {Electromagnetic Probes of Hot and Dense\\
         Hadronic Matter}}
\vskip 0.1in
Jan-e Alam$^{a,b},$\footnote{presented at the DAE symposium on Nuclear Physics,
Mumbai, December 1998.},
Sourav Sarkar$^a$,  Pradip Roy$^c$ and 
Bikash Sinha$^{a,c}$

{\it a) Variable Energy Cyclotron Centre,
     1/AF Bidhan Nagar\\
     Calcutta 700 064 India}\\

{\it b) Physics Department, Kyoto University, Kyoto 606-8502, Japan}\\

{\it c) Saha Institute of Nuclear Physics,
           1/AF Bidhan Nagar\\
           Calcutta 700 064 India}\\
\end{center}

%\date{\today}
\parindent=20pt
\vskip 0.2in
\begin{abstract}
The formalism of 
photon and dilepton productions in a thermal bath are discussed
in the frame work of finite temperature field theory. 
The emission rate has been expressed in terms of the discontinuities 
or imaginary part of the photon self energy in the thermal bath. 
The photon and dilepton emissions have been computed by taking into 
account the in-medium modifications of the particles 
appearing in the internal loop of the self energy diagram  
using a phenomenological effective lagrangian approach. The results
has been compared with the photon spectra measured by  
WA 98 collaboration at CERN SPS energies.
\end{abstract}

\section*{I. Introduction}

\begin{figure}
\centerline{\psfig{figure=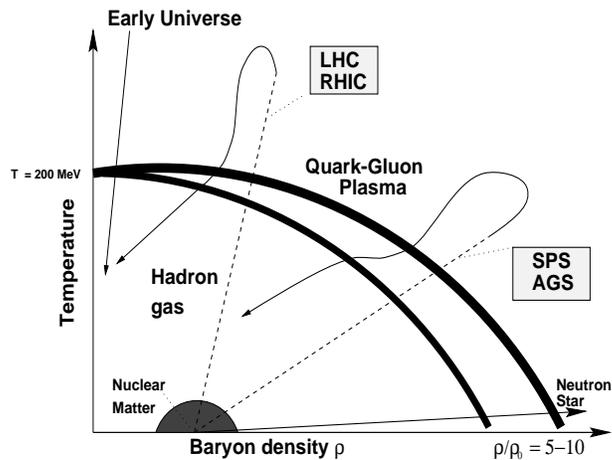,height=6cm,width=8cm}}
\caption{ QCD phase diagram.}
\label{QCDfig}
\end{figure}

Numerical simulations of the QCD (Quantum Chromodynamics) equation of state 
on the lattice predict that at very high density and/or temperature 
hadronic matter undergoes a phase transition to Quark Gluon Plasma 
(QGP)~\cite{ukawa}(see Fig.~\ref{QCDfig}). 
One expects that ultrarelativistic heavy ion 
collisions might create conditions conducive for the formation and study 
of QGP. Various model calculations have been
performed to look for observable signatures of this state of matter.
However, among various signatures of QGP, photons and
dileptons are known to be advantageous,
primarily  so  because, electromagnetic interaction could
lead to detectable signal. However, it is weak enough to let the
produced particles (real photons and dileptons) escape the system
without further interaction and thus carrying the information 
of the constituents and their momentum distribution 
in the thermal bath.

The disadvantage with photons is the substantial background from 
various processes (thermal and non-thermal)~\cite{janepr}. Among these, 
the contribution from hard QCD processes is well understood in the 
framework of perturbative QCD and the yield from hadronic decays 
e. g. $\pi^0\,\ra\,\gamma\,\gamma$ can be accounted for by invariant mass 
analysis. However, photons from the thermalised hadronic gas pose a more
difficult task to disentangle. Therefore it is very important to estimate
photons from hot and dense hadronic gas along with the possible modification
of the hadronic properties.

%%%%%%%%%%%% Figure 1 %%%%%%%%%%%%%%%%%%%%%
%%%%%%%%%%%%%%%%%%%%%%%%%%%%%%%%%%%%%%%%%%%

The importance of the electromagnetic probes for  the  thermodynamic
state  of the evolving matter was first proposed by Feinberg in 
 1976~\cite{feinberg}. While for most purposes one can 
calculate  the  emission rates in a classical framework, Feinberg
showed  that  the  emission  rates  can   be   related   to   the
electromagnetic   current  current  correlation  functions  in  a
thermalised  system  in  a  quantum  picture   and,   more
importantly,  in  a nonperturbative manner. 
Generally the production of a particle which interact weakly 
with the constituents of the thermal bath ( the constituents 
may interact strongly among themselves, we need not give
any explicit form of their coupling strength)
can always be expressed
in terms of the discontinuities or imaginary parts of the self
energies of that particle~\cite{ruuskanen,bellac} which never thermalizes. 
We should, therefore look at the
connection between the  electromagnetic  emission  rates  (real
photons and lepton pairs, which correspond to virtual photons)
and  the photon spectral function ( which is 
connected with the discontinuities in self energies) 
in a thermal system~\cite{weldon90}, 
which in turn connected to the hadronic electromagnetic current current
correlation tensor~\cite{mclerran} through Maxwell equations. 
We will show below that 
the photon emission rate can be obtained from the dilepton
(which originate from a virtual photon) emission rate with little effort.

In section II we present the 
general formalism for the dilepton and photon production rate 
from a thermal bath. In section III we calculate the medium 
modifications of hadrons. Section IV is devoted to discuss
the results of our calculations. In section V we present
a summary and discussion.

\section*{II. Electromagnetic Probes - Formulation}

We begin our discussion with the dilepton production rate.
In the following we will follow the  work of
Weldon~\cite{weldon90} for the derivation of dilepton
emission rate.
Let $e_0\,J_\mu\,A^\mu$ be the interaction between the photon and the particles
in the heat bath, then to lowest order in the electromagnetic coupling, the 
$S$ matrix element for the transition  
$\mid I\rangle\,\rightarrow\,\mid H;l^+l^-\rangle$ is given by
\be
S_{HI}=e_0\,\int\,\langle\,H;l^+l^-\,\mid\,J_\mu\,A^\mu(x)\,\mid\,I\rangle\,
d^4x\,e^{iq\cdot x}
\ee
where $\mid I\rangle$ is the initial state corresponding to the two
incoming nuclei, $\mid H;l^+l^-\rangle$ is the final state 
which corresponds to a lepton pair plus anything (Hadronic), 
the parameter $e_0$ is the
bare (unrenormalised) charge and $q=p_1\,+\,p_2$ is the 
four momentum of the lepton pair.
Since we assume that the lepton pair do not interact
with the emitting system, the matrix element can be factorised 
in the following way,
\begin{equation}
\langle\,H;l^+l^-\mid\,J_\mu\,A^\mu(x)\,\mid\,I\rangle=
\langle H\mid\,A^\mu(x)\mid\,I\rangle\langle\,l^+l^-\mid\,J_\mu\,\mid\,0\rangle
\end{equation}
Putting the explicit form of the current $J^\mu$ in terms of the
Dirac spinors we obtain
\begin{equation}
S_{HI}=e_0\,\frac{\bar{u}(p_1)\gamma_\mu\,v(p_2)}{V\sqrt{2E_12E_2}}\int
d^4x\,e^{iq\cdot x}\langle H\mid\,A^\mu(x)\mid I\rangle
\end{equation}
Therefore dilepton multiplicity from a thermal system is obtained as,
(by summing over the final states and averaging over the initial 
states with a weight factor $Z(\beta)^{-1}\,e^{-\beta\,E_I}$)
\begin{equation}
N=\frac{1}{Z(\beta)}\sum_I\,\sum_H\,\mid S_{HI}\,\mid^2\,e^{-\beta\,E_I}
\end{equation}
where $E_I$ is the total energy in the initial state.
After some algebra $N$ can be written in a compact form as follows,
\begin{equation}
N=e_0^2\,L^{\mu\nu}\,H_{\mu\nu} \frac{d^3p_1}{(2\pi)^32E_1}
\,\frac{d^3p_2}{(2\pi)^32E_2}
\end{equation}
where $L_{\mu\nu}$ is the leptonic tensor given by
\begin{eqnarray}
L^{\mu\nu}&=&\frac{1}{4}\,
\sum_{spins}\,\bar{u}(p_1)\gamma^\mu\,v(p_2)\bar{v}(p_2)
\gamma^\nu\,u(p_1)\nonumber\\
&=&\,p_1^\mu\,p_2^\nu+p_2^\mu\,p_2^\nu-\frac{q^2}{2}\,g^{\mn}
\end{eqnarray}
and $H_{\mu\nu}$ is the photon tensor
\be
H_{\mn}=\frac{1}{Z(\beta)}\,e^{-\beta q_0}\sum_H\,\int d^4x\,d^4y\,
e^{iq\cdot (x-y)}
\,\langle\,H\mid\,A_\mu(x)\,A_\nu(y)\mid\,H\rangle\,e^{-\beta\,E_H}
\ee
to obtain the above equation we have used 
the resolution of identity $1=\sum_I\,\mid I\rangle\langle I\mid$, and
the energy conservation equation $E_I=E_H+q_0$ 
where $q_0$ is the energy of the lepton pair and $E_H$ is the energy
of the rest of the system produced after collision.
Using the translational invariance of the matrix element we can write
\begin{equation}
H_{\mu\nu}=2\pi\,\Omega\,e^{-\beta\,q_0}\rho^{\mu\nu}
\end{equation}
where $\Omega$ is the four volume of the system and $\rho^{\mu\nu}$
is the spectral function of the photon in the heat bath,
\be
\rho_{\mn}(\vec q,q_0)=\frac{1}{2\pi\,Z(\beta)}\int d^4x\,e^{iq\cdot x}\sum_H
\,\langle\,H\mid\,A_\mu(x)\,A_\nu(0)\mid\,H\rangle\,e^{-\beta\,E_H}
\label{spectral1}
\ee
The rate of dilepton production per unit volume ($N/\Omega$) is given by,
\be
\frac{dN}{d^4x}=2\pi\,e_0^2 L_{\mn}\,e^{-\beta q_0}\rho^{\mn}(q)
\frac{d^3p_1}{(2\pi)^32E_1}\,\frac{d^3p_2}{(2\pi)^32E_2}
\label{funda}
\ee
This result which expresses the dilepton emission rate 
in terms of the spectral function of the photon in the medium 
is a fundamental result.
At zero temperature ($\beta\,\rightarrow\,\infty$) the only state which
enters in the spectral function ($\rho(s)$) is the vacuum and the 
spectral function can be expressed in terms of
the imaginary part of Greens function (i.e the discontinuity along
the real axis) $\rho(s)=-\frac{1}{\pi}\,{\s Im}\,G(s)$~\cite{lsbrown}.

Unlike at zero temperature,
in the case of non-zero temperature the time ordered ($D^{\mu\nu}$)
and the Feynman propagator ($D^{\mn}_F$) are
different. The relation between them is given by~\cite{nieves},  
\be
{\s Im}\,D^{\mn}=(1+2f_{BE}){\s Im}\,D^{\mn}_F
\label{toptof}
\ee
where $f_{BE}$ is the thermal distribution for photons.
Now the time ordered Green's function can be expressed in terms of the
spectral function as follows
\be 
D^{\mn}(q_0,\vec q)=\int_{-\infty}^{\infty}\,d\alpha\,
\left[\frac{\rho^{\mn}(\alpha,\vec q)}{q^0-\alpha+i\epsilon}
-\frac{\rho^{\nu\mu}(\alpha,-\vec q)}{q^0+\alpha-i\epsilon}\right]
\label{top}
\ee
Using the KMS relation one can show that 
\be
\rho^{\mn}(q_0,\vec q)=e^{\beta\,q_0}\rho^{\nu\mu}(-q_0,-\vec q)
\label{kms}
\ee
 substituting Eq.~(\ref{kms}) in Eq.~(\ref{top}) and using 
Eq.~(\ref{toptof})
we obtain
\be
\rho^{\mn}(q_0,\vec q)=
\frac{-1}{\pi}\,(1+f_{BE}(q_0))
{\s Im}\,D^{\mn}_F(q_0,\vec q)
\label{spectral}
\ee
The above equation implies that to evaluate the spectral function
at $T\neq 0$ we need to know the imaginary part of the Feynman propagator.
It is important to note that above expression for spectral function
reduces its vacuum value as $\beta\ra\,\infty$.
The Feynman propagator can be expressed in terms of the proper self 
energy as
\be
D^{\mn}_F=-\left[\,\frac{Z_3\,A^{\mn}}{q^2-\Pi_T}+
\frac{Z_3\,B^{\mn}}{q^2-\Pi_L}\right] + (\zeta-1)\frac{q_\mu\,q_\nu}{q^4}
\label{selfe}
\ee
where $Z_3$ is the wave function renormalization constant at zero 
temperature,  $A^{\mn}$ and $B^{\mn}$ are transverse and longitudinal 
projection tensors respectively.
The presence of the parameter $\zeta$ indicates the gauge 
dependence of the propagator. Although the gauge dependence cancels out
in the calculation of physical quantities,
one should, however, be careful when extracting
physical quantities from the propagator directly, especially in the
non-abelian gauge theory.

Using Eqs.~(\ref{funda}), (\ref{spectral}) and (\ref{selfe})  we 
get 
\be
\frac{dN}{d^4x}=2\,e^2\,L_{\mn}(A^{\mn}\rho_T\,+\,B^{\mn}\,\rho_L)
\frac{d^3p_1}{(2\pi)^32E_1}\,\frac{d^3p_2}{(2\pi)^32E_2}\,
f_{BE}(q_0)
\label{dilrate}
\ee
where $e^2=Z_3\,e_0^2$, is the physical charge of the electron and 
\be
\rho_k=\frac{{\s Im}\,\Pi_k}{(q^2-Re\,\Pi)^2+{\s Im}\,\Pi_k^2} 
\ee 

The emission rate of dilepton can also be expressed 
in terms of the electromagnetic
current-current correlation functions~\cite{mclerran}.
Denoting the hadronic part of the electromagnetic  current operator
by  $J^h_\mu$,  the leptonic part by $J^l_\nu$ and the free photon
propagator by $D^{\mn}$, we have the matrix element  for  this
transition :
\be
S_{HI}=\langle\,H;l^+\,l^-\,\mid\int d^4x d^4y
J^h_{\mu}(x)D^{\mn}\,J^l_{\nu}(y)\mid I\rangle
\ee
As in the earlier case the leptonic part of the current 
can be easily factored out. 
Writing  the  Fourier transform of photon propagator and squaring
the matrix elements,
one obtains, for the rate of dilepton
production, 
\be
dR=e_0^2\,L^{\mn}\,W_{\mn}\,\frac{1}{q^4}\,
\frac{d^3\,p_1}{(2\pi)^3\,E_1}
\frac{d^3\,p_2}{(2\pi)^3\,E_2}
\label{dnd4x}
\ee
where $W^{\mu\nu}(q)$ is just the Fourier transform of the thermal 
expectation value
of the real time electromagnetic current-current correlation function :
\be
W^{\mu\nu}=\int d^4x e^{-iqx}\sum_H\,
\langle H\,\mid J^{\mu}(x)J^{\nu}(0)\mid\,H\,\rangle\,
\frac{e^{-\beta\,E_H}}{Z}
\label{correl}
\ee
The subtleties of the thermal averaging have been elucidated earlier.
It is thus readily seen (eq.~\ref{dnd4x}) that the dilepton data
yields considerable information about the thermal state 
of the hadronic system; at least the full tensor structure
of $W^{\mu\nu}$ can in principle be determined.
The most important point to realize here is
that the analysis is essentially nonperturbative.

At this point one may wonder what is the connection between 
the electromagnetic current-current correlation function and the spectral
function? The connection can be expressed in a straight forward way
by substituting $J_\mu$ and $J_\nu$ 
using Maxwell equation ($\partial_\alpha\partial^\alpha\,A^\mu
-\partial^\mu\,(\partial_\alpha\,A^\alpha)=j^\mu$) in Eq.~(\ref{correl})
to obtain,
\be
e_0^2\,W^{\mn}=2\pi\,(q^2\,g^{\mu\alpha}-q^\mu\,q^\alpha)\,
\rho_{\alpha\beta}\,(q^2\,g^{\beta\nu}-q^\beta\,q^\nu)
\ee

In the literature most of the dilepton production rate from
a thermal system are calculated with the approximation
$\Pi_T=\Pi_L$ which gives the spectral function as
\be
\rho^{\mn}(q_0,\vec q)=(-g^{\mn}+q^\mu\,q^\nu/q^2)
\frac{-1}{\pi}\,(1+f_{BE}(q_0))\,{\s Im}\,\left[\frac{1}{q^2-\Pi_L}\right]
\label{restspec}
\ee
Since ${\s Im}\,\Pi_k$ and ${\s Re}\,\Pi_k$ are proportional
to $\alpha$ (the fine structure constant) they are very small
for most of our practical purposes. Therefore $\rho_k$ is given by
\be
\rho_k=\frac{{\s Im}\,\Pi_k}{q^4}
\label{approx}
\ee
Using Eqs.~(\ref{dilrate}), (\ref{restspec}), (\ref{approx})
and the following result
\begin{eqnarray}
\int\,\prod_{i=1,2}\,\frac{d^3p_i}{(2\pi)^32E_i}
\delta^4(p_1+p_2-q)\,L^{\mn}(p_1,p_2)&=&\frac{1}{(2\pi)^6}\frac{2\pi}{3}
(1+\frac{2m^2}{q^2})\nonumber\\
&&\times\sqrt{1-\frac{4m^2}{q^2}}C^{\mn}
\end{eqnarray}
where $C^{\mn}=q^\mu\,q^\nu-q^2\,g^{\mn}$,
we get,
\be
\frac{dR}{d^4q}=\frac{\alpha}{12\pi^4\,q^2}(1+\frac{2m^2}{q^2})
\sqrt{1-\frac{4m^2}{q^2}}{\s Im}\Pi^\mu_\mu\,f_{BE}(q_0)
\label{usedrate}
\ee
This is the familiar results most widely used for dilepton emission
rate~\cite{bellac}. 

\begin{figure}
\centerline{\psfig{figure=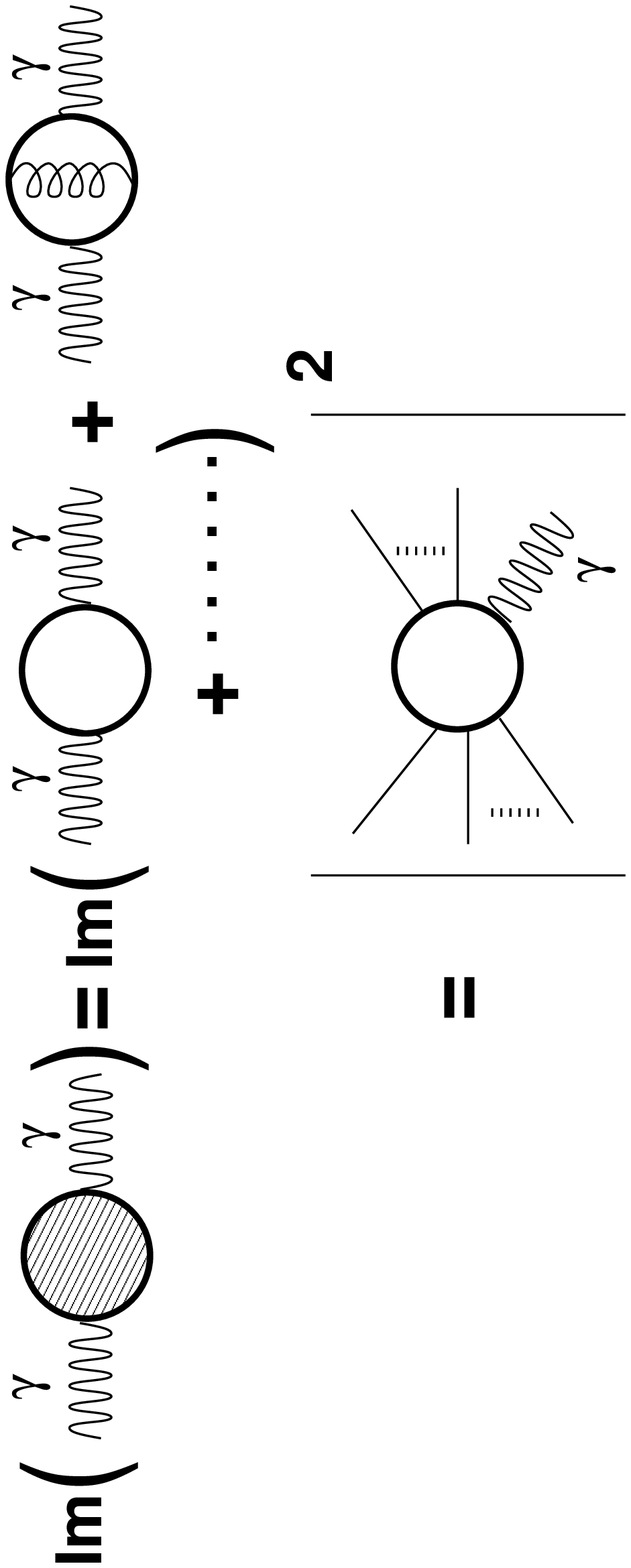,width=15cm,height=6cm,angle=-90}}
\caption{Optical Theorem in Quantum Field Theory}
\label{opt}
\end{figure}

To obtain real photon emission rate per unit volume 
($dR$) from a system under thermal 
equilibrium we note that the dilepton emission rate differs from the
photon emission rate in the following way. The factor $e^2\,\,L_{\mn}\,
\frac{1}{q^4}$ is the product of the electromagnetic vertex 
$\gamma^\ast\,\ra\,l^+\,l^-$, the leptonic current involving Dirac spinors
and the square of the photon propagator should be replaced by the factor 
$\sum\,\epsilon_\mu\,\epsilon_\nu (=-g_{\mn})$ for the real (on-shell)
photon. Finally the phase space factor 
$d^3p_1/[(2\pi)^32\,E_1]\,d^3p_2/[(2\pi)^32E_2]$
should be replaced by, $d^3q/[(2\pi)^32q_0]$ to obtain
\be
dR=-\frac{1}{(2\pi)^3}\,g_{\mn}\,W^{\mn}\,\frac{d^3q}{q_0}
\ee
The current-current correlation function is related to the photon self
energy as~\cite{bellac,beres,kapusta} 
\be
W_{\mn}=2\,f_{BE}(q_0)\,{\s Im}\,\Pi_{\mn}
\ee
Therefore,
\be
q_0\frac{dR}{d^3q}=-2\frac{1}{(2\pi)^3}\,{\s Im}\Pi^\mu_\mu\,f_{BE}(q_0) 
\ee
The above equation can also be obtained directly from eq.~\ref{usedrate}.
The emission rate given above is correct up to order $e^2$ in 
electromagnetic interaction but exact, in principle, to all order
in strong interaction. But for all practical purposes 
one is able to evaluate up to a finite order of loop expansion.
Now it is clear from the above results 
that to evaluate photon and dilepton emission rate from a 
thermal system we need to evaluate the 
imaginary part of the photon self energy.
The Cutkosky rules at finite temperature~\cite{kobes,adas,gelis}
gives a systematic procedure to calculate the imaginary part of a 
Feynman diagram. The Cutkosky rule expresses the 
imaginary part of the $n$-loop amplitude in terms of physical 
amplitude of lower order ($n-1$ loop or lower). This is shown schematically
in Fig.~\ref{opt}.  When the imaginary part of the self energy is calculated
up to and including $L$ order loops where $L$ satisfies $x\,+\,y\,<\,L\,+\,1$
then we obtain the photon emission rate for the reaction; $x$ particles
$\ra$ $y$ particles $+\,\gamma$ and the above formalism becomes 
equivalent to the relativistic kinetic theory formalism~\cite{gale}. 

%%%%%%%%%%%%% Fig.2 %%%%%%%%%%%%%%%%%%%%%%%%%%%%%
%%%%%%%%%%%%% Fig.2 %%%%%%%%%%%%%%%%%%%%%%%%%%%%%
For a reaction $1\,+\,2\,\ra\,3\,+\gamma$ the photon (of energy E)
emission rate is given by~\cite{sourav},
\begin{eqnarray}
E\frac{dR}{d^3p}&=&\frac{N}{16(2\pi)^7E}\,\int_{(m_1+m_2)^2}^{\infty}
\,ds\,\int_{t_{\s {min}}}^{t_{\s {max}}}\,dt\,|{\cal M}|^2\,
\int\,dE_1\nonumber\\
&&\times\int\,dE_2\frac{f(E_1)\,f(E_2)\left[1+f(E_3)\right]}{\sqrt{aE_2^2+
2bE_2+c}}
\label{rktp}
\end{eqnarray}
where
\begin{eqnarray}
a&=&-(s+t-m_2^2-m_3^2)^2\nonumber\\
b&=&E_1(s+t-m_2^2-m_3^2)(m_2^2-t)+E[(s+t-m_2^2-m_3^2)(s-m_1^2-m_2^2)\nonumber\\
&&-2m_1^2(m_2^2-t)]\nonumber\\
c&=&-E_1^2(m_2^2-t)^2-2E_1E[2m_2^2(s+t-m_2^2-m_3^2)-(m_2^2-t)(s-m_1^2-m_2^2)]
\nonumber\\
&&-E^2[(s-m_1^2-m_2^2)^2-4m_1^2m_2^2]-(s+t-m_2^2-m_3^2)(m_2^2-t)\nonumber\\
&&\times(s-m_1^2-m_2^2)
+m_2^2(s+t-m_2^2-m_3^2)^2+m_1^2(m_2^2-t)^2\nonumber\\
E_{1{\s {min}}}&=&\frac{(s+t-m_2^2-m_3^2)}{4E}+\frac{Em_1^2}{s+t-m_2^2-m_3^2}
\nonumber\\
E_{2{\s {min}}}&=&\frac{Em_2^2}{m_2^2-t}+\frac{m_2^2-t}{4E}\nonumber\\
E_{2{\s {max}}}&=&-\frac{b}{a}+\frac{\sqrt{b^2-ac}}{a}\nonumber
\end{eqnarray}

In a similar way the dilepton emission rate for a reaction
$a\,\bar a\,\ra\,l^+\,l^-$ can be obtained as
\begin{eqnarray}
\frac{dN}{d^4xd^3qdM}&=&\int {d^3p_a\over 2E_a(2\pi)^3}f(p_a)
\int {d^3p_{\bar a}\over 2E_{\bar a}(2\pi)^3}f(p_{\bar a})
\int {d^3p_1\over 2E_1(2\pi)^3}
\int {d^3p_2\over 2E_2(2\pi)^3}\nonumber\\
&&\mid M\mid_{a\bar a\rightarrow l^+l^-}^2
(2\pi)^4\delta(\vec p_a+\vec p_{\bar a}-\vec p_1-\vec p_2)
\delta(E_a+E_{\bar a}-E_1-E_2)\nonumber\\
&&\delta(p_a+p_{\bar a})
\delta(M-E_a-E_{\bar a}) 
\end{eqnarray}
where
$M^2=(p_a+p_{\bar a})^2$ is the invariant mass
and $f(p_a)$ is the
occupation probability in the momentum space.
The Pauli blocking of the lepton pair in the final state has been
neglected in the above equation. 

\section*{III. Finite Temperature Properties}

To calculate the effective mass and decay widths of vector mesons 
we begin with the following $VNN$ (Vector - Nucleon - Nucleon) Lagrangian 
density:

\begin{equation}
{\cal L}_{VNN} = g_{VNN}\,\left({\bar N}\gamma_{\mu}
\tau^a N{V}_{a}^{\mu} - \frac{\kappa_V}{2M}{\bar N}
\sigma_{\mu \nu}\tau^a N\partial^{\nu}V_{a}^{\mu}\right),
\label{lag1}
\end{equation}
where $V_a^{\mu} = \{\omega^{\mu},{\vec {\rho}}^{\mu}\}$,
$M$ is the free nucleon mass, $N$ is the nucleon field
and $\tau_a=\{1,{\vec {\tau}}\}$.
The values of the coupling constants $g_{VNN}$ and $\kappa_V$ will 
be specified later.  With the above Lagrangian we proceed to calculate the 
$\rho$-self energy. 

\begin{equation}
\Pi_{\mu \nu} = -2ig_{VNN}^2\,\int\,
\frac{d^4{p}}{(2\pi)^4}\,K_{\mu \nu}(p,k),
\label{pimunu}
\end{equation}
with,
\begin{equation}
K_{\mn} = \frac{{\rm Tr}\left[\Gamma_{\mu}(k)\,(p\!\!\!/+M^{*})
\Gamma_{\nu}(-k)\,(p\!\!\!/-k\!\!\!/+M^{*})\right]}{(p^2-M^{* 2})
((p-k)^2-M^{*2})},
\label{smn}
\end{equation}
The vertex $\Gamma_{\mu}(k)$ is calculated by using the
Lagrangian of Eq.~(\ref{lag1}) and is given by

\begin{equation}
\Gamma_{\mu}(k) = \gamma_{\mu} + \frac{i\kappa_V}{2M}\sigma_{\mu \alpha}
k^{\alpha}.
\label{vertex}
\end{equation}
Here $M^{\ast}$ is the in-medium (effective) mass of the nucleon at finite 
temperature which we calculate using the Mean-Field Theory (MFT)~\cite{vol16}.
The value of $M^{\ast}$ can be found by solving the following 
self consistent equation:

\begin{equation}
M^{\ast} = M -\frac{4g_s^2}{m_s^2}\,\int\,\frac{d^3{\bf p}}{(2\pi)^3}\,
\frac{M^{\ast}}{({\bf p}^2+M^{\ast 2})^{1/2}}\,\left[f_N(T)+f_{\bar N}(T)\,\right],
\label{MN}
\end{equation}
where $f_N(T)(f_{\bar N}(T))$ is the Fermi-Dirac distribution for 
the nucleon (antinucleon), $m_s$ is mass of the neutral scalar meson
($\sigma$) field, and, the nucleon interacts via the exchange of 
isoscalar meson with coupling constant $g_s$.  Since the exact 
solution of the field equations in QHD is untenable, these are solved in 
mean field approximation. In a mean field approximation one
replaces the field operators by their ground state expectation
values which are classical quantities; this 
renders the field equations exactly solvable. 

The vacuum part of the $\rho$ self energy arises due to its interaction 
with the nucleons in the Dirac sea. This is calculated using dimensional 
regularization scheme (see~\cite{pradip} for details).

In a hot system of particles, there is a thermal distribution of real particles
(on shell) which can participate in the absorption and emission process in
addition to the exchange of virtual particles. The interaction of the rho 
with the onshell nucleons, present in the thermal bath contributes to the 
medium dependent part of the $\rho$-self energy.  This is calculated from
Eq.(\ref{pimunu}) using imaginary time formalism as,

\begin{eqnarray}
{\s {Re}}\Pi_{\s {med}}(\omega,{\bf k}\,\ra\,0)& = &
-\frac{16g_{VNN}^2}{\pi^2}\,\int\,\frac{p^2\,dp}{\omega_p\,
(e^{\beta\,\omega_p}+1)(4\omega_p^2-\omega^2)}\nonumber\\
&&\times\,\left[\frac{1}{3}(2p^2+3M^{\ast 2})+\omega^2\left\{M^{\ast}
(\frac{\kappa_V}
{2M})\right.\right.\nonumber\\
&&+\left.\left.\,\frac{1}{3}(\frac{\kappa_V}{2M})^2(p^2+3M^{\ast 2})\right\}
\right]
\end{eqnarray}
where, $\omega_p^2=p^2+M^{\ast 2}$.

\section*{IV. Results}

   In this section we present the results of our calculation of the effective
masses and decay widths of vector mesons and its effect on electromagnetic
probes. 
For our calculations of $\rho$ and $\omega$~mesons 
effective masses we have used the following values of the coupling constants 
and masses~\cite{bonn}: $\kappa_{\rho} = 6.1,~g_{\rho NN}^2/4\pi = 0.55, 
m_s$= 550 MeV, $m_{\rho} = 770$ MeV, $M = 938$ MeV, $g_s^2/4\pi = 9.3$,
$g_{\omega NN}^2/4\pi = 20$, and $\kappa_{\omega}$ = 0.
In Fig.~3 we show the variation of the ratio of effective mass to free 
mass for hadrons as a function of temperature. 
The expected trend based on Brown - Rho scaling is also shown. 
Firstly, we do not observe any global scaling behaviour. Secondly, 
in our case the effective mass as a function of temperature falls
at a slower rate.  

%%%%%%%%%%%%%%%%%%%%%%Figure 3 %%%%%%%%%%%%%%%%%%%%%%%%%%%%%%%%%%%%%
\begin{figure}
\centerline{\psfig{figure=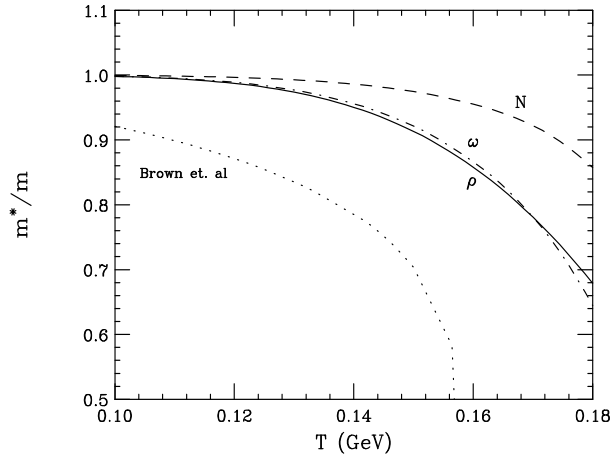,height=6cm,width=8cm}}
\caption{
Ratio of effective mass to free space mass of hadrons as a 
function of temperature
$T$.}
\label{scaling}
\end{figure}
%%%%%%%%%%%%%%%%%%%%%%%%%%%%%%%%%%%%%%%%%%%%%%%%%%%%%%%%%%%%%%%

Many authors have investigated the issue of temperature dependence of
hadronic masses within different models over the past several years.
Hatsuda and collaborators~\cite{furn,adami,hatsuda} and Brown~\cite{Brown}
showed that the use of QCD sum rules at finite temperature results in
a temperature dependence of the $q \bar q$ condensate culminating
in the following behaviour of the $\rho$-mass:
\be
\frac{m_{\rho}^{\ast}}{m_{\rho}} = \left(1- \frac{T^2}{T_{\chi}^ 2}\right)^
{1/6}
\ee
where $T_{\chi}$ is the critical temperature for chiral phase
transition. Brown and Rho~\cite{rho} also showed that the requirement 
of chiral symmetry yields an approximate scaling relation between 
various effective hadron masses,
\be
\frac{M^{\ast}}{M}\approx\frac{m_{\rho}^{\ast}}{m_{\rho}}\approx
\frac{m_{\omega}^{\ast}}{m_{\omega}}\approx\frac{f_{\pi}^{\ast}}{f_{\pi}}
\ee
These calculations show a dropping of hadronic mass with temperature. 
Calculations with non-linear $\sigma$-model, however,  
predict the opposite trend \cite{abijit}. 

%%%%%%%%%%%%%%%%%%%% Figure 4 %%%%%%%%%%%%%%%%%%%%%%%%%%% 
\begin{figure}
\centerline{\psfig{figure=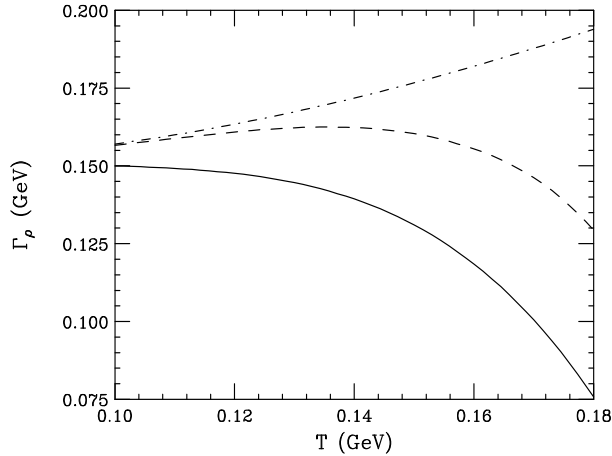,height=6cm,width=8cm}}
\caption{
Rho decay width ($\Gamma_{\rho}$) as a function of temperature ($T$).
Dashed and solid lines show calculations of rho width with effective mass
due to nucleon loop, with and without BE. Dot dashed line represents
the same but with effective mass due to pion loop. 
}
\label{rowidth}
\end{figure}
%%%%%%%%%%%%%%%%%%%%%%%%%%%%%%%%%%%%%%%%%%%%%%%%%%%%%%%%%%%

The variation of the in-medium decay width ($\Gamma_{\rho}$) of the rho meson
with temperature is shown in Fig.~(\ref{rowidth}). As discussed earlier,
the effective mass of the rho decreases as a result of $N\bar N$ 
polarisation. This reduces the phase space available for the rho. Hence, 
we observe a rapid decrease in the rho meson width with temperature 
(solid line). However, the presence of pions in the medium would
cause an enhancement of the decay width through induced emission. Thus
when Bose Enhancement (BE) of the pions is taken into account 
the rho decay width is seen to fall less rapidly (dashed line);
such a behaviour is observed quite clearly in ~\cite{abhee}.
For the sake of completeness, we also show the variation of rho width in
the case where the rho mass changes due to $\pi \pi$ loop.
In this case, since the rho mass increases (though only marginally), the width 
increases (dot-dashed line).

Now we present our results on photon emission rates from a
hot hadronic gas. As discussed earlier, the variation of hadronic decay
widths and masses will affect the photon spectra. The relevant reactions
of photon production are $\pi \pi\,\ra\,\rho\,\gamma, \pi\,\rho\,\ra
\pi\,\gamma, \pi\,\pi\,\ra\,\eta\,\gamma$, and $\pi\,\eta\,\ra\,\eta\,\gamma$
with all possible isospin combinations. Since the lifetimes of the rho and
omega mesons are comparable to the strong interaction time scales,
the decays $\rho\,\ra\,\pi\,\pi\,\gamma$ and $\omega\,\ra\,\pi^0\,\gamma$
are also included. 
The effect of finite resonance width of the rho meson in the photon 
production cross-sections has been taken into account through the
propagator. 
%%%%%%%%%%%%%%%%%%%%%%% Figure 5 %%%%%%%%%%%%%%%%%%%%%%%%
\begin{figure}
\centerline{\psfig{figure=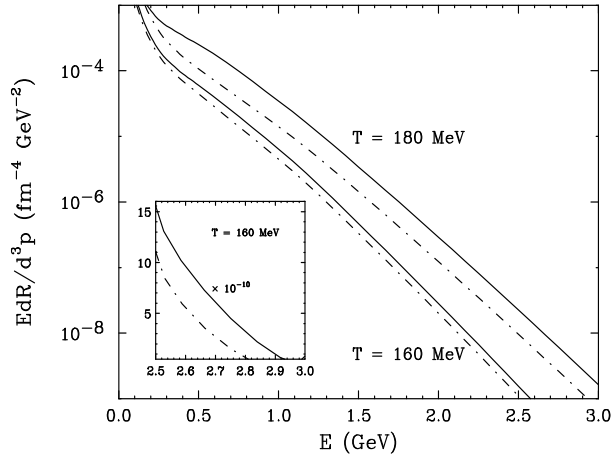,height=6cm,width=8cm}}
\caption{
Total photon emission rate from hot hadronic matter
as a function of photon energy at $T$=160, 180 MeV. 
The solid and dot-dashed lines show results
with and without in-medium effects respectively.
Inset: Total photon emission rate is plotted in linear
scale as a function of photon energy in the kinematic 
window, $E_{\gamma}=$ 2.5 to 3.0 GeV at $T=$180 MeV.
}
\label{totphot}
\end{figure}
%%%%%%%%%%%%%%%%%%%%%%%%%%%%%%%%%%%%%%%%%%%%%%%%%%%%%%%%%

In Fig.~(\ref{totphot}) we display the total rate of emission 
of photons from hot
hadronic gas including all hadronic reactions and decays of vector
mesons. At $T=$180 MeV, the photon emission rate with
finite temperature effects is a factor of $\sim$ 3 higher 
than the rate calculated without medium effects.
At $T=160$ MeV the medium effects are small compared to the previous
case.
  
\begin{figure}
\centerline{\psfig{figure=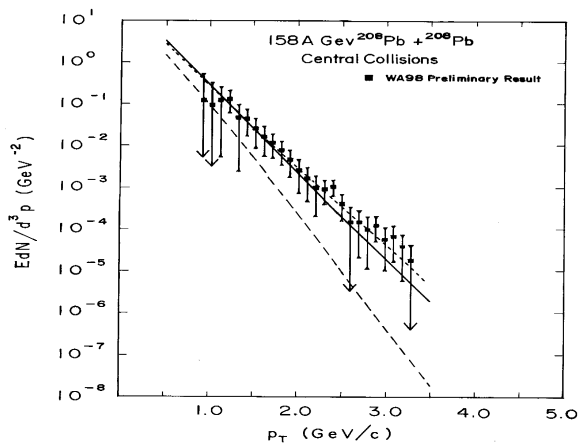,height=6cm,width=8cm}}
\caption{Total thermal photon yield in Pb + Pb central collisions
at 158 GeV per nucleon at CERN SPS. The long-dash line shows the results
when the system is formed in the QGP phase 
with initial temperature $T_i=180$ MeV
at $\tau_i=1$ fm/c. The critical temperature for phase transition is
taken as 160 MeV. The solid
(short-dash) line indicates photon spectra  
when hadronic matter formed in the initial state at $T_i=230$ MeV
($T_i=270$ MeV) at $\tau_i=1$ fm/c 
with (without) medium effects on hadronic masses and decay widths.
}
\label{wa98}
\end{figure}

The transverse momentum distribution of photon has been obtained
by convoluting the production rate with the space time
dynamics of the system. The boost invariant hydrodynamical 
model of Bjorken~\cite{bjorken} has been used for the space
time evolution of the system with appropriate modification
due to the shift in hadronic masses~\cite{spjs}. 
In Fig.~\ref{wa98} we compare our results of transverse momentum
distribution of photons  with the preliminary results of WA98
Collaboration~\cite{wa98con}. 
The experimental data represents the photon spectra
from Pb + Pb collisions at 158 GeV per nucleon at CERN
SPS energies. The transverse momentum distribution of photons originating
from the `hadronic scenario' (matter formed in the hadronic
phase) with (solid line) and without(short-dash line) 
medium modifications of vector mesons outshine the photons 
from the `QGP scenario' (matter formed in the QGP phase, indicated 
by long-dashed line) for the entire range of $p_T$.
Although photons from `hadronic scenario' with medium
effects on vector mesons shine less bright than those from
without medium effects for $p_T>2$ GeV, it is not possible
to distinguish clearly between one or the other on the basis of the 
experimental data.  
%%%%%%%%%%%%%% Fig.6 %%%%%%%%%%%%%%%%%%%%%%%%%%%%%
%%%%%%%%%% End of Fig. 6 %%%%%%%%%%%%%%%%%%%%%%%%%%%%%%%%

So far we have studied the effects of in-medium hadronic masses
and decay widths on the photon spectra, where, 
the latter is found to have negligible effects. However, the modification in the
rho decay width due to BE, arising from the induced emission of 
pions in the thermal medium, plays  a crucial role in  low mass 
lepton pair production. It has been shown
in Ref.~~\cite{weldon} that this effect arises naturally in a calculation 
based on finite temperature field theory.

\begin{figure}
\centerline{\psfig{figure=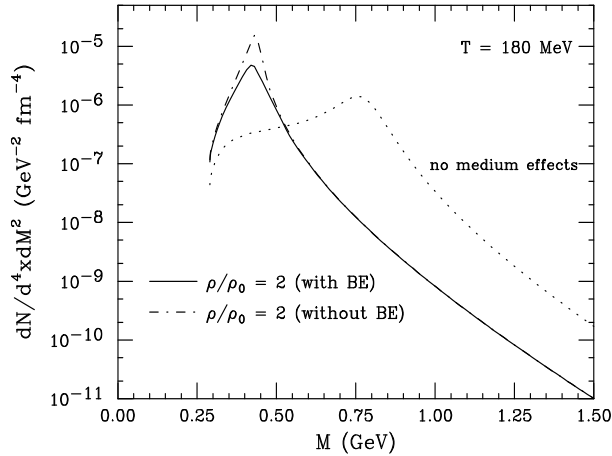,height=6cm,width=8cm}}
\caption{
Dilepton emission rates as a function of invariant mass 
at $T$ = 180 MeV and twice nuclear matter density. Solid and dotdashed 
lines show the results with and without BE effects respectively. The dotted
line shows the result when no medium effect is considered.
}
\label{ratedil}
\end{figure}

For the sake of illustration,
we consider the effect of thermal nucleon loop on the lepton pair 
production from pion annihilation via (rho mediated) vector meson 
dominance. In the work of Li, Ko and Brown~\cite{li}, the observed
enhancement on the low mass dilepton yield was attributed solely to the
the matter induced mass modification of the
meson, while neglecting the BE effect. 
In Fig.~(\ref{ratedil}) the dilepton emission rate is plotted 
as a function of invariant mass at a temperature  $T$=180 MeV. 
It is clear from the figure that the inclusion of BE effects 
reduces the yield by increasing the decay width. Quantitatively,
a suppression by a factor of 3 is observed at the invariant mass,
$M=m_{\rho}^{\ast}$.

%%%%%%%%%%%%%% Figure 7 %%%%%%%%%%%%%%%%%%%%%%%%%%%%%
%%%%%%%%%%%%%%%%%%%%%%%%%%%%%%%%%%%%%%%%%%%
%\vskip 0.2cm

\section*{V. Summary and Discussions}

We have discuss the general formalism to evaluate the emission of
electromagnetic ejectiles from a thermal bath. This formalism has been
applied to evaluate the photon and dilepton emission rate from a hot
and dense hadronic system. 
We have calculated the effective masses and decay widths
of vector mesons propagating in a hot medium. We have seen that the 
mass of rho meson decreases substantially due to its interactions with
nucleonic excitations and it increases only marginally ($\sim$ 10-15 MeV) 
due to $\rho-\pi$ interactions.  The overall decrease in the effective 
mass is due to fluctuations in the Dirac sea of nucleons with mass 
$M^{\ast}$ and the in-medium contribution increases very little 
($\sim$ 5 -- 10 MeV) from its free space value. 
The omega mass drops at $T=$180
MeV by about 250 MeV from its free space value. 

We have evaluated the rate of photon emission from a hadronic gas of $\pi,
\rho, \omega$ and $\eta$ mesons. It is seen that the photon rate increases by a
factor $\sim$ 3  due to the inclusion of in-medium masses at $T=$180 MeV.
We observe that the inclusion of in-medium decay widths in vector meson 
propagator has negligible effects on photon emission rates, although 
the effect of in-medium decay widths with BE is substantial for the 
dilepton yield.   

We have compared the experimental data on photon spectra from  
Pb + Pb collisions at energies 158 GeV per nucleon 
with different initial conditions.  We observe that
photons from hadronic scenario dominates over the photons from
QGP scenario for the entire $p_T$ domain. 
But in the hadronic scenario the photon spectra evaluated with
and without in-medium properties of vector mesons describe 
these data reasonably well. Hence the 
transverse photon spectra at present do not allow us to decide between
an in-medium dropping mass and a free mass scenario. 
Considering the experimental uncertainty, it is not possible
to state, which one, between the two is compatible with the data.
Experimental data with better statistics could possibly
distinguish among various scenarios.

\noindent{{\bf Acknowledgement}:
We thank B. Dutta-Roy, H. Gutbrod and V. Manko for useful discussions.
One of us (JA) is grateful to Japan Society for Promotion 
of Science for financial support and Physics Department, Kyoto
University where part of this manuscript was written.}\\

%%%%%%%%%%%%%%%%%%%%%%%%%%%%%%%%%%%%%%%%%%%%%%%%%%%%%%%%%%%%%%%%%%%%%%%%%


\begin{thebibliography}{59}

\bibitem{ukawa} A. Ukawa, Nucl. Phys. A{\bf 638} 339c (1998).

\bibitem{janepr} J. Alam, S. Raha and B. Sinha, Phys. Rep. {\bf 273}
243 (1996).

\bibitem{feinberg} E. L. Feinberg, Nuovo Cim. {\bf 34A} (1976)39.

\bibitem{ruuskanen} P. V. Ruuskanen in
 Particle Production in Highly Excited Matter,
 ed. H. H.  Gutbrod, (Plenum Press, New York, 1992)

 \bibitem{bellac} M. Le Bellac, Thermal Field Theory,
Quantum Field Theory, Cambridge
University Press, 1996.

 \bibitem{weldon90} H. A. Weldon, Phys. Rev. {\bf D42} (1990)2384.

 \bibitem{mclerran} L. McLerran and T. Toimela, Phys. Rev. {\bf D31}
 (1985)545.

 \bibitem{lsbrown} L. S. Brown, Quantum Field Theory, Cambridge
University Press, 1995.

 \bibitem{nieves} J. F. Nieves, Phys. Rev. {\bf D 42} (1990) 4123.

 \bibitem{beres} V. B. Berestetskii, E. M. Lifshitz and L. P.
Pitaevskii, Quantum Electrodynamics, Pregamon Press, 1982.

 \bibitem{kapusta} C. Gale and J. Kapusta, Nucl. Phys. 
{\bf B357}, (1991) 65.

 \bibitem{kobes} R. Kobes and G. Semenoff, Nucl Phys. {\bf B260} 
(1985) 714; {\bf B272}, (1986) 329. 

 \bibitem{adas} A. Das, Finite Temperature Field Theory,
(World Scientific, Singapore, 1997).

 \bibitem{gelis} F. Gelis, Nucl. Phys. {\bf B508}, (1997) 483. 

 \bibitem{gale} C. Gale and J. I. Kapusta, Phys. Rev. {\bf C35}, 
(1987)2107. 

 \bibitem{sourav} S. Sarkar et al, Nucl. Phys. {\bf A634}, (1998)206. 

\bibitem{vol16} B. D. Serot and J. D. Walecka, Advances in Nuclear Physics
Vol. 16 Plenum Press, New York 1986. 

 \bibitem{pradip} P. Roy, S. Sarkar, J. Alam and 
B. Sinha , nucl-th/9803052.

\bibitem{bonn} R. Machleidt, K. Holinde, and Ch. ELster, Phys. Rep.
{\bf 149} 1 (1987).

\bibitem{furn} R. J. Furnstahl and T. Hatsuda, Phys. Rev. {\bf D42} 
1744 (1990). 

\bibitem{adami} C. Adami, T. Hatsuda, and I. Zahed, Phys. Rev. {\bf D43}
921 (1991).

\bibitem{hatsuda} T. Hatsuda, Nucl. Phys. {\bf A544} 27c (1992).

\bibitem{Brown} G. E. Brown, Nucl. Phys. {\bf A522} 397c (1991).

\bibitem{rho} G. E. Brown and M. Rho, Phys. Rev. Lett. {\bf 66}
2720 (1991).

\bibitem{abijit} A. Bhattacharyya, J. Alam, S. Raha and B. Sinha,
Int. J. Mod. Phys. {\bf A} {\bf 12} (1997) 5639.

\bibitem{abhee} A. K. Dutt-Majumdar, J. Alam, B. Dutta-Roy, and B. Sinha,
Phys. Lett. {\bf B378} 35 (1996).

\bibitem{bjorken} J. D. Bjorken, Phys. Rev. D {\bf 27} (1983)140.

\bibitem{spjs} S. Sarkar, P. Roy, J. Alam and B. Sinha, nucl-th/9812006.

\bibitem{wa98con} V. Manko, Int. Nucl. Phys. Conf. (INPC-98), 
August `98,  Paris, France. 

\bibitem{weldon} H. A. Weldon,  Phys. Rev. D {\bf 28} (1983)2007.

\bibitem{li} G. Q. Li, C. M. Ko and G. E. Brown, Nucl. Phys. A {\bf 606}
(1996)568.

\end{thebibliography}
\end{document}